\documentclass[aps,pra,twocolumn,showpacs,superscriptaddress,groupedaddress]{revtex4-1}

\usepackage{amsmath,hyperref,braket,amssymb,natbib,graphicx,float,amsthm}

\begin{document}

\title{Quantum Hypergraph States in Continuous Variables}
\author{Darren W. Moore}
\email{darren.moore@upol.cz}
\affiliation{Department of Optics, Palack\'{y} University, 17. listopadu 1192/12, 771 46 Olomouc, Czech Republic}

\begin{abstract}
The measurement based, or one-way, model of quantum computation for continuous variables uses a highly entangled state called a cluster state to accomplish the task of computing. Cluster states that are universal for computation are a subset of a class of states called graph states. These states are Gaussian states and therefore require that the homodyne detection (Gaussian measurement) scheme is supplemented with a non-Gaussian measurement for universal computation, a significant experimental challenge. Here we define a new non-Gaussian class of states based on hypergraphs which satisfy the requirements of the Lloyd-Braunstein criteria while restricted to a Gaussian measurement strategy. Our main result is to show that, taking advantage of the intrinsic multimode nonlinearity, a hypergraph consisting of 3-edges can be used to apply a three-mode operation to an input three-mode state. As a special case, this technique can be used to apply the cubic phase gate to a single mode. 
\end{abstract}

\maketitle

\section{Introduction}%%%%%%%%%%INTRODUCTION

Recent years have seen a notable development of interest in technologies capable of taking advantage of the nonlinear aspects of quantum mechanics especially in optomechanics~\cite{thompson2008strong,karuza2012tunable,paraiso2015positionsquared,kalaee2016design,brawley2016nonlinear,gieseler2013thermal,kiesel2014cavity,fonseca2016nonlinear,rashid2016experimental,siler2017thermally,ricci2017optically}, whether electromechanical, membrane in the middle or levitated systems, or in optical systems~\cite{ourjoumtsev2007increasing,marek2012deterministic,marshall2015repeat,miyata2016implementation}. In parallel there is a strong thread of development of quantum information processing with continuous variables (CV)~\cite{lloyd2003quantum,menicucci2006universal,weedbrook2012gaussian}. These research trajectories are complementary in that the most interesting quantum information processing tasks for continuous variables tend to require non-Gaussian operations~\cite{fiurasek2002gaussian,gu2009quantum,niset2009nogo,arzani2017polynomial,walschaers2018tailoring} which are implemented via dynamics incorporating nonlinearities in the quadrature/mode operators. Simultaneously the development of quantum information processing tasks which require such nonlinear dynamics provides a stimulus to develop theoretical insight into nonlinear quantum mechanics~\cite{braunstein2001otimal,fiurasek2002gaussian,eisert2002distilling,giedke2002characterization,olivares2003teleportation,niset2009nogo,adesso2009optimal,paternostro2009violations,magnin2010strong,jabbour2015interconversion,zhuang2018resource,albarelli2018resource} and experimental progress towards implementing such tasks~\cite{miwa2009demonstration,ukai2011demonstration,yokoyama2013ultra,yoshikawa2016invited,miyata2016implementation}. A clear example of the linear/nonlinear distinction is afforded in quantum computing wherein universality, defined as the ability to simulate an arbitrary Hamiltonian up to arbitrary accuracy, requires a non-Gaussian operation as well as access to the class of Gaussian operations. This translates directly into a distinction between linear and nonlinear operations. 

Analogously, discrete variable (DV) systems require, in addition to Clifford operations, access to a non-Clifford operation. In both cases, the straightforward approach is to find an interaction that takes the set of states outside the limiting Gaussian/Clifford classes, or in the case of measurement based quantum computation (MBQC) non-Clifford/non-Gaussian projective measurements. In CV systems this is particularly difficult as engineering nonlinear Hamiltonians remains a challenge. Alternative strategies have been developed both for DV and CV, often with a clear thematic link between the two. Thus, in DV computing, one has the concept of magic states~\cite{bravyi2005universal}, a set of ancilla states which in addition to Clifford operations permit universal quantum computation. The analogy in CV systems is found with the cubic phase state playing the role of the magic state~\cite{menicucci2014fault}, such that the measurement strategy is composed of only homodyne detection (Gaussian measurements). The idea can be recast in terms of seeding a cluster state with the requisite ancilla state~\cite{houhou2018unconditional}. Note that, since CV computation requires a discrete encoding for fault-tolerance, there are nonlinear states forming the encoding that allow the measurement strategy to remain Gaussian~\cite{gottesman2001encoding,baragiola2019all}. 

Some recent developments in quantum information have lead to research into discrete variable hypergraph states~\cite{rossi2013quantum,qu2013encoding} along with their entanglement and symmetry properties~\cite{miller2016hierarchy,devakul2018universal,miller2018latent}. These special symmetry properties allow a properly prepared hypergraph to implement universal measurement based quantum computation (MBQC) while retaining a measurement strategy that draws only from the class of Pauli operators, although this is not an exhaustive description of systems with such a property~\cite{kissinger2019universal}.

In what follows we introduce the class of CV hypergraph states. Unlike cluster states, such states have a highly nonlinear preparation which obviates the need for nonlinear measurements in order to implement nonlinear gate operations. We will demonstrate that hypergraphs with minimal nonlinearity, produced by cubic interactions, are sufficient for for universal quantum computing with CV while retaining a Gaussian measurement strategy. To accomplish this, we design a protocol taking advantage of the inherent multimode nonlinearity to apply a three-mode operation on an arbitrary three-mode input state. As a special case, this operation is used to implement the single mode cubic phase gate.

\section{CV Hypergraph States}%%%%%%%%%%CV HYPERGRAPH STATES

A hypergraph is a pair $(V,E)$ where the set $V$ contains the $n$ vertices of the hypergraph and the elements of $E=\{(v_{i_1},v_{i_2},\dots,v_{i_k})|v_{i_j}\in V\}$ are $k$-tuples describing which subsets of $V$ are connected by a hyperedge and $i=1,\dots,$ labels each hyperedge. A useful classification is that of $k$-uniformity in which a $k$-uniform hypergraph contains only edges of order $k$.

Let $\ket{0}_p$ denote an eigenstate of momentum with eigenvalue zero. Despite their idealised nature, these states form a useful mathematical representation easily mapped to the formalism of hypergraphs. They will represent the elements of the set $V$. To define the edges we introduce the entangling operator $CZ_{i_1,\dots,i_k}=e^{iq_{i_1}\dots q_{i_k}}$, with $q$ the canonical position operator, which forms a generalisation of the controlled phase gates to $k-1$ control qumodes. Note that $CZ_i$ denotes the familiar displacement in momentum $Z(-1)=e^{iq}$ and $CZ_{i,j}$ the standard $CZ$ gate $e^{iq_iq_j}$. Furthermore, observe that this generalisation is still symmetric in the control qumodes i.e. any $k-1$ collection of $k$ modes may constitute the set of control qumodes. Finally we may define the $k$-uniform hypergraph states as follows
\begin{equation}
\ket{g_k}=\prod_{e_k\in E_k}CZ_{e_k}\ket{0}_p^{\otimes n}\,,
\end{equation}
where $E_k\subset E$ denotes the set of hyperedges of order $k$ and $e_k\in E_k$ is a hyperedge. Observe that the states $\ket{g_1}$ are a collection of uncorrelated eigenstates of momentum and $\ket{g_2}$ are the standard graph states, of which cluster states form a subset.

The definition of fully general hypergraphs involves a product over all $k\le n$:
\begin{equation}
\ket{g_{\le n}}=\prod_{k=1}^n\prod_{e_k\in E_k}CZ_{e_k}\ket{0}_p^{\otimes n}\,.\label{Hypergraph}
\end{equation}
These include all possible hyperedges up to order $n$. The construction of an $n$ vertex hypergraph may then include $n$-body interactions among the qumodes. To illustrate consider a simple linear four-vertex hypergraph with a 3-edge between vertices 1, 2 and 3, and a 2-edge between vertices 3 and 4. Such a graph represents the state
$e^{iq_1q_2q_3}e^{iq_3q_4}\ket{0}_p^{\otimes4}$.

\subsection{Stabilisers and Nullifiers}

As an introduction to stabilisers and nullifiers for hypergraph states consider the same for graph states. First recall the position and momentum displacement operators $X(s)=e^{isp}$, with $p$ the momentum conjugate to $q$, and $Z(s)=e^{-isq}$. The stabilisers are the set of operators
\begin{equation}
K^{(2)}_i(s)=X_i(s)\bigotimes_{j\in\mathcal{N}(i)}Z_j(s)=X_i(s)\bigotimes_{j\in\mathcal{N}(i)}CZ_j(s)\,,
\end{equation}
where $\mathcal{N}(i)=\{j|(i,j)\in E_2)\}$ is the neighbourhood of the vertex $i$. The nullifiers follow from the definition of stabiliser as $K_i^{(2)}(s)\ket{g_2}=\ket{g_2}\forall s\in\mathbb{R}$. They are
\begin{equation}
H^{(2)}_i=p_i-\sum_{j\in\mathcal{N}(i)}q_j\,,
\end{equation}
and they have the property that $H_i\ket{g_2}=0$.

Extending to $k$-uniform hypergraph states, the stabilisers are written
\begin{equation}
K^{(k)}_i(s)=X_i(s)\bigotimes_{e_{k-1}\in\mathcal{N}(i)}CZ_{e_{k-1}}\,,
\end{equation}
where now the neighbourhood is extended to the hyperedges connected to vertex $i$ by $\mathcal{N}(i)=\{e_{k-1}|e_{k-1}\cup\{i\}=e_k\in E_k\}$. Similarly the nullifiers can be extended, deriving directly from the stabilisers, as follows
\begin{equation}
H^{(k)}_i=p_i-\sum_{e_{k-1}\in\mathcal{N}(i)}q_{i_1}\dots q_{i_{k-1}}\,.
\end{equation}

Finally, these operators are extended to the full hypergraph states by including all possible types of hyperedges. The stabilisers are
\begin{equation}
K^{(\le n)}_i(s)=X_i(s)\prod_{k=1}^n\bigotimes_{e_{k-1}\in\mathcal{N}(i)}CZ_{e_{k-1}}\,,
\end{equation}
and the nullifiers
\begin{equation}
H^{(\le n)}_i=p_i-\sum_{k=1}^n\sum_{e_{k-1}\in\mathcal{N}(i)}q_{i_1}\dots q_{i_{k-1}}\,.
\end{equation}

It is easy to see that these relations are true by considering that if $X$ is a stabiliser for $\ket{\psi}$, then $UXU^\dagger$ is a stabiliser for $U\ket{\psi}$. Then consider $X(s)$ acting on a momentum eigenstate and the effect of the unitary operators acting to produce the hypergraph state. It is easy to verify that the nullifiers satisfy the property $[H^{(\le n)}_i,H^{(\le n)}_j]=0$. Furthermore, any $H$ in the nullifier space ($H\ket{g_{\le n}}=0$) can be expressed as a linear combination of the core nullifiers defined above, i.e. $H=\sum_jc_jH^{(\le n)}_j$.

To illustrate these concepts consider vertex 3 of the same example presented above. Then the stabiliser has the form $K_3(s)=e^{isp_3}e^{iq_1q_2}e^{iq_4}$ and the nullifier $H_3=p_3-q_1q_2-q_4$. 

\section{Multimode Nonlinear Operations}%%%%%%%%%%MULTIMODE NONLINEAR OPERATIONS

\begin{figure}[h]
\includegraphics[width=\columnwidth]{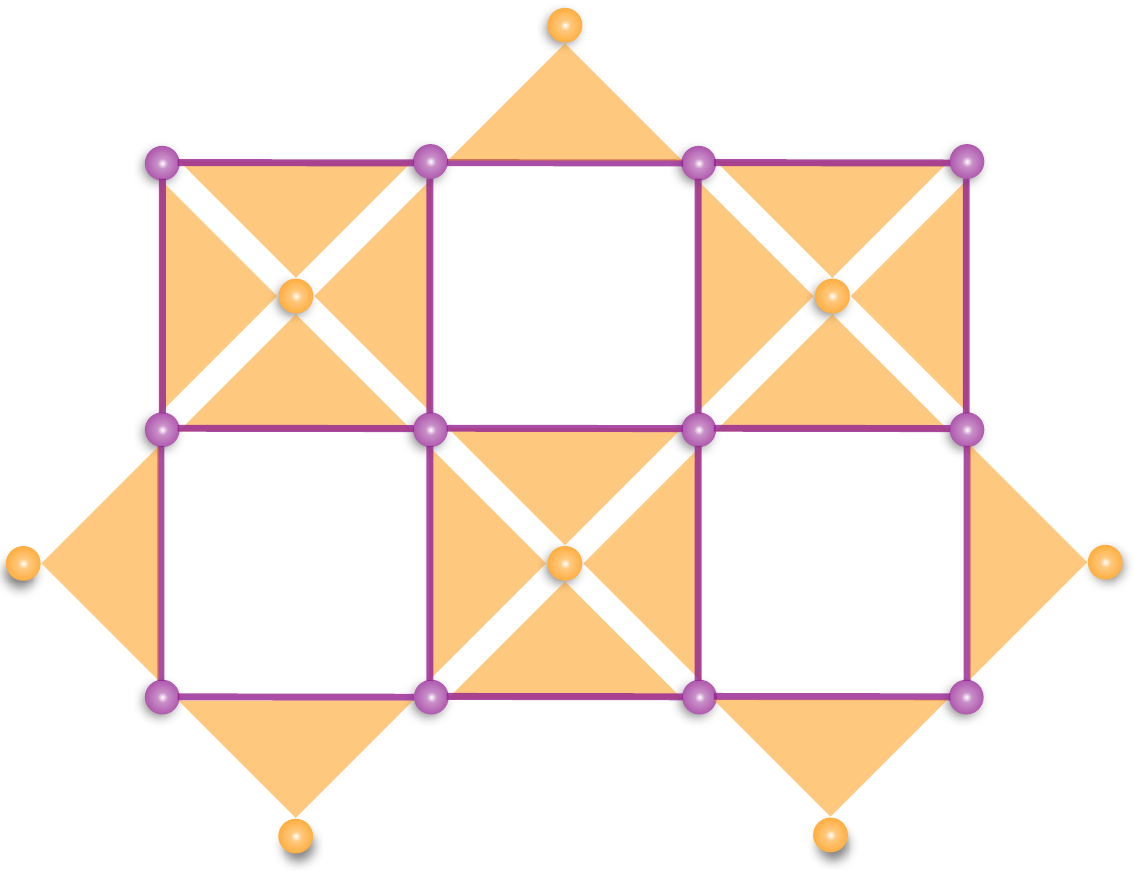}
\caption{The repeating lattice of 3-edges forming the 3-cluster hypergraph. A hypergraph consisting only of 3-edges can be converted to a standard cluster state via $q$ measurements on selected nodes.  Orange nodes are those to be measured in the preparation of a standard cluster state. The purple nodes are those remaining connected to the graph post-measurement, connected via the newly generated 2-edges portrayed as solid purple lines. The result is a 2D lattice cluster state.}\label{3graph}
\end{figure}

The power of graph states inheres in their structure, and for hypergraph states this power finds its form in the nonlinearity dispersed among multiple modes. The focus of this article will be on a hypergraph built out of 3-edges which we refer to as a 3-cluster state (Fig.~\ref{3graph}). This means that the nonlocal nonlinearity will have the cubic form $q_iq_jq_k$~\cite{frattini20173wave}. In fact this cubic form implies the nullifiers of the 3-cluster will take the form $p_i-q_jq_k$, suggestive of the nonlinear squeezing resource required for adaptively implementing the cubic phase state via measurement~\cite{miyata2016implementation,moore2019estimation}. This 3-edge hypergraph bears much similarity to the Union Jack state of Ref~\cite{miller2016hierarchy} and has the ability, through Gaussian measurements, to teleport a 3-edge onto a new set of modes not previously sharing a 3-edge. Gaussian measurements on hypergraphs have the effect of reducing the order of the hyperedges connected to the measured node by one. This is detailed further in Lemmas 1 and 2 in the Appendix.

%The `Gaussian universality' property will be explored for the case of quantum hypergraph states involving 3-edges, analogous with the Union Jack state of Ref~\cite{miller2016hierarchy}. The hypergraph state we consider will be referred to as a 3-cluster state (Fig.~\ref{3graph}). The demonstration that a Gaussian measurement strategy is sufficient will consist of two parts: firstly we show that arbitrary Gaussian operations can be accomplished using $q$ and $p$ measurements on a 3-cluster, and secondly that arrangements can be made for a cubic phase gate to be applied to an arbitrary input state.

\begin{figure}[h]
\includegraphics[width=\columnwidth]{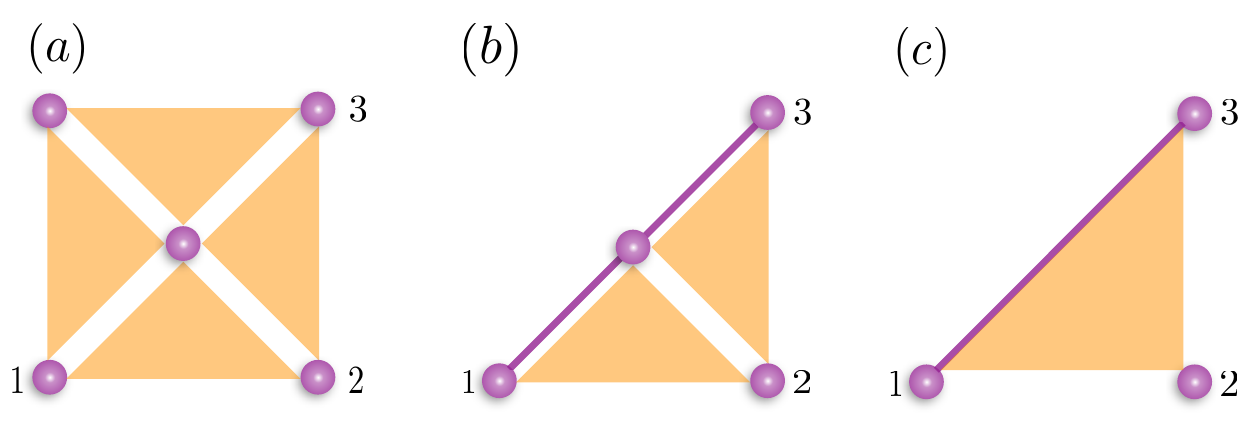}
\caption{(a) A cell of the 3-cluster. Vertices are purple nodes and 3-edges are orange triangles. (b) The upper left vertex is measured in the computational basis $q$, leaving the configuration shown. Then, a measurement in the $p$ basis is performed on the remaining central vertex to give configuration (c). Solid lines denote standard 2-edges, generated by $CZ$ and the triangles denote 3-edges. The configuration in (c) is equivalent to a 3-edge with Gaussian by-products. See the text for a fuller description of the correlations involved.}\label{nonG}
\end{figure}

\textbf{Theorem}: \textit{A 3-uniform hypergraph (3-graph) under a Gaussian measurement scheme is sufficient to generate a 3-edge between 3 vertices not previously sharing such a hyperedge.}

\textit{Proof}: Consider a 3-uniform hypergraph state with a repeating structure comprised of a cell consisting of a central vertex and four vertices forming an enclosing square alternating with squares lacking a central vertex. The 3-edges are applied on the four triangles formed between the central vertex and its enclosure (see Fig.~\ref{3graph}). Take a cell of the lattice involving a central vertex and measure $q$ on the upper left corner (see Fig.~\ref{nonG}). Then, using Lemma 1, the 3-edges are modified to 2-edges (standard $CZ$ edges). Next, a $p$ measurement is made on the central vertex (Lemma 2). This produces a state described by the following expressions. (The Gaussian by-products from the first measurement are omitted for simplicity.)
\begin{align}
&\int dxe^{-imx}e^{ix(q_1q_2+q_2q_3+q_1+q_3)}\ket{0}_p^{\otimes3}=\\&
=\int dxdye^{-imx}e^{ix(q_1q_2+q_2q_3+q_3)}\ket{x}_p\ket{y}_q\ket{0}_p\\
&=\int dxdye^{-imx}e^{ixq_3}e^{ixyq_1}\ket{x}_p\ket{y}_q\ket{xy}_p\\
&=\int dxdye^{-imx}e^{ixq_3}e^{ip_3q_1}\ket{x}_p\ket{y}_q\ket{xy}_p\\
&=e^{ip_3q_1}F_1CZ_{13}Z_1(m)\int dxdy\ket{x}_q\ket{y}_q\ket{xy}_p\\
&=e^{ip_3q_1}F_1CZ_{13}Z_1(m)CZ_{123}\ket{0}^{\otimes3}_p
\end{align}
\qed

What this shows is that it is possible to generate a nonlinear three-mode operation between a set of vertices that were not previously connected by a 3-edge. That is, the cell in Fig.~\ref{nonG} is a device allowing 3-edges to be teleported around the hypergraph state. The calculation also makes clear a difference from the standard paradigm of MBQC, in that the byproducts are not necessarily local. This follows from the properties of measurements on hyperedges explored in Lemmas 1 and 2 i.e. measurement of an edge reduces its order by one. However, for the case of a 3-cluster we note that the byproducts are indeed always Gaussian. As a special case, aimed towards universal computing, this technique can be adapted to perform the cubic phase gate on some input state. 

\textbf{Corollory}: \textit{The cubic phase gate can be implemented on an input state connected to a 3-cluster using only Gaussian measurements.}

\textit{Proof}: Consider the following state, generated by Gaussian operations on an arbitrary input state and two Gaussian ancillas:
\begin{align}
&\int dx\psi(x)\ket{x}_q\ket{\gamma x}_q\ket{x}_q=\\
&=\int dx\psi(x)\ket{x}_qF_3^\dagger S(\gamma) e^{-iq_2q_3}\ket{x}_q\ket{0}_p\\
&=F_3^\dagger S(\gamma)e^{-iq_2q_3}F_2^\dagger e^{-iq_1q_2}\int dx\psi(x)\ket{x}_q\ket{0}_p\ket{0}_p\\
&=F_3^\dagger S(\gamma)e^{-iq_2q_3}F_2^\dagger e^{-iq_1q_2}\ket{\psi}\ket{0}_p\ket{0}_p
\end{align}
The final line consists of a collection of Gaussian operations on a product state. Applying a 3-edge to this state, using the Theorem, followed by $p$ measurements on the ancilla modes produces a cubic phase gate applied to the input $\ket{\psi}$. 
\begin{multline}
e^{iq_1q_2q_3}\int dx\psi(x)\ket{x}_q\ket{\gamma x}_q\ket{x}_q=\\
\int dx\psi(x)e^{i\gamma x^3}\ket{x}_q\ket{\gamma x}_q\ket{x}_q
\end{multline}
\begin{align}
{}_{p_2}\bra{n}{}_{p_1}\bra{m}&\rightarrow\int dx\psi(x)e^{i\gamma x^3}e^{-im\gamma x}e^{-inx}\ket{x}_q\\
&=X(\gamma m+n)e^{i\gamma q^3}\ket{\psi}
\end{align}
\qed

By preparing an appropriate three mode ancilla the cubic phase gate can be applied to an arbitrary input state using only Gaussian measurements. The strength of the nonlinearity $\gamma$ in the cubic gate is determined by the initial squeezing applied to one of the ancilla modes. What remains to satisfy the criteria for universal quantum computation with continuous variables is to ensure that all Gaussian operations are accessible. 

It is already known that the standard 2-edge cluster state is universal for Gaussian operations. Knowing that Gaussian measurements on the 3-cluster convert 3-edges into 2-edges (see Appendix) we can find a measurement strategy that converts regions of the 3-cluster into standard cluster states.  

 %\textbf{Theorem 1}: \textit{A 3-uniform hypergraph (3-graph) under a Gaussian measurement scheme is sufficient to perform universal Gaussian computation. In particular the 2D lattice cluster state used in standard MBQC is reproducible deterministically (disregarding Gaussian corrections) from a 3-graph.}

Consider the 3-uniform hypergraph state with the alternating geometry from (Fig.~\ref{3graph}). By Lemma 1 (see Appendix) if measurements of $q$ are made on the central vertices then they are disconnected from the graph. Furthermore, the 3-edges are modified to 2-edges (standard $CZ$ edges) in every direction, forming a square lattice of 2-edges. This is indeed the required cluster state, ignoring Gaussian corrections.

The 2D cluster state thus generated is universal for Gaussian operations under Gaussian measurements. It is proper to give some attention to the effects of the Gaussian byproducts of the process, as they may affect the quality of the resulting cluster state. According to Lemma 1, each reduced hyperedge gains a weight. For 3-edges the outcome of the measurements produces a weighted graph state wherein the weights on each 2-edge are determined by the outcome of the measurement. This is easily interpreted as a measurement-dependent squeezing $m$ of the remaining modes (see Lemma 1). As seen in the Lemma, on one hand there is a Gaussian byproduct acting on the resultant state. On the other, the ideal momentum eigenstates absorb the effect of the squeezing directly applied to the cluster vertex. In a realistic scenario using momentum-squeezed states, this is unlikely to occur and the squeezing may be detrimental to the cluster, especially since fault-tolerance relies on high levels of squeezing~\cite{menicucci2014fault}. There are three cases to consider; $m>1$ which anti-squeezes the desired momentum-squeezed mode,  $m<1$ which enhances the squeezing of the mode and $m=0$ in which the identity operator is applied. A consequence of this process is that the entire lattice will not be uniformly squeezed, since different measurements will have independent outcomes. 

%Given the nullifier definition for a 3-edge, $H_i^{(3)}=p_i-q_jq_k$, it is to be expected that the 3-cluster has the potential to implement a cubic phase gate, $e^{i\gamma q^3}$. The nullifier represents an infinite squeezing in a nonlinear quadrature which is precisely analogous to the resource required for the cubic phase gate~\cite{miyata2016implementation}. 

\section{Conclusion}%%%%%%%%%%CONCLUSION

To review, a 3-cluster state is generated as a hypergraph state consisting of 3-edges arranged as in Fig.~\ref{3graph}. The Theorem indicates that the multimode nonlinearity can be used to apply a nonlocal, nonlinear operation on a set of input modes. This apparatus can be co-opted into a degenerate form in which the cubic nonlinearity is condensed into a single mode, producing the cubic phase gate and leaving the two other input modes as ancillas. Finally, a region of a 3-cluster can be converted into a standard cluster state in order to take advantage of its inherent Gaussian universality. With these procedures in place the Lloyd-Braustein criteria for computing with continuous variables are satisfied, making the 3-cluster a candidate for universal quantum computing.

We have defined a new class of CV states, generalising the notion of graph states to non-Gaussian hypergraph states. The class of 3-hypergraphs can act as a tool for performing a multimode nonlinear operation on arbitrary input states and satisfies the typical strategy for showing universality for continuous variables. Furthermore this is all accomplished while restricted to a Gaussian measurement strategy. Hypergraph states exemplify nonlinear phase states for multimode systems and higher orders may also prove to have the structure necessary to carry out interesting nonlinear and multimode operations. Moreover under Gaussian measurements higher order hyperedges do not reduce to Gaussian byproducts, meaning that such higher order structures will blend the capacities of various nonlinearities.  

\begin{acknowledgments}
DM thanks Alessandro Ferraro and Radim Filip for their helpful discussions. The author has received national funding from the MEYS of the Czech Republic (Project No. 8C18003) under Grant Agreement No. 731473 within the QUANTERA ERA-NET Cofund in Quantum Technologies implemented within the European Union's Horizon 2020 Programme (ProjectTheBlinQC) as well as support from the Czech Science Foundation under project 19-17765S from the Development Project of Faculty of Science, Palack\'{y} University.
\end{acknowledgments}

\appendix

\section{Shaping Hypergraph States}\label{appendix}

There are two basic ways to shape hypergraph states using Gaussian measurements: position measurements or momentum measurements. Lemma 1 addresses position measurements and Lemma 2 addresses momentum measurements.

\textbf{Lemma 1}: \textit{A $q$ measurement on a vertex of a hypergraph state disconnects the measured vertex from the hypergraph, converting any hyperedges on the affected vertex from $k$-hyperedges to $(k-1)$-hyperedges along with a Gaussian by-product.}

\textit{Proof}: Consider a generic hypergraph state, as in Eq.~(\ref{Hypergraph}). Perform a measurement of $q$ on an arbitrary mode $j$ with result $m$. For notational brevity, define $E^k_+\subset E$ to be the set of $e_k\in E_k$ such that $e_k\in E^k_+\Rightarrow j\in e_k$. Then define $E^k_-\subset E$ to be the set of $e_k\in E_k$ such that $j\notin e_k$. Without loss of generality let $j$ be the $k$th vertex for $e_k\in E_+$. Now the measurement has the following effect:
\begin{widetext}
\begin{align}
\ket{m}_{q,j}\braket{m|g_{\le n}}&=\prod_{k=1}^n\prod_{e_k\in E_k}\ket{m}_{q,j}\bra{m}CZ_{e_k}\ket{0}_p^{\otimes n}\\
&=\prod_k\prod_{e_k\in E^k_-}CZ_{e_k}\prod_{e_k\in E^k_+}\ket{m}_{q,j}\bra{m}CZ_{e_k}\ket{0}_p^{\otimes n}\\
&=\prod_k\prod_{e_k\in E^k_-}CZ_{e_k}\prod_{e_k\in E^k_+}\ket{m}_{q,j}\bra{m}e^{iq_{i_1}\dots q_{i_k}}\ket{0}_p^{\otimes n}\\
&=\prod_k\prod_{e_k\in E^k_-}CZ_{e_k}\prod_{e_{k-1}\in E^{k-1}_+}e^{imq_{i_1}\dots q_{i_{k-1}}}\ket{0}_p^{\otimes n-1}\ket{m}_{q_k}\,,
\end{align}
\end{widetext}
Now, noting that $m$ and $\ket{m}_{q_k}$ are an eigenvalue and eigenstate of $q_k$,
\begin{widetext}
\begin{align}
\ket{m}_{q,j}\braket{m|g_{\le n}}&=\prod_k\prod_{e_k\in E^k_-}CZ_{e_k}\prod_{e_{k-1}\in E^{k-1}_+}F^\dagger_kF_ke^{iq_{i_1}\dots q_{i_k}}F^\dagger_k\ket{0}_p^{\otimes n-1}\ket{m}_{p_k}\\
&=\prod_k\prod_{e_k\in E^k_-}CZ_{e_k}\prod_{e_{k-1}\in E^{k-1}_+}F^\dagger_ke^{iq_{i_1}\dots q_{i_{k-1}}p_k}e^{imq_k}\ket{0}_p^{\otimes n-1}\ket{0}_{p_k}\,.
\end{align}
\end{widetext}
Commute the displacement through the graph operator. Furthermore, for a squeezing operator defined by $S(s)=e^{-\frac i2\ln s(qp+pq)}$, the effect on position eigenstates is $S^\dagger\ket{q}_q=\ket{\frac qs}_q$, where the squeezing is in momentum (position) for $s>1$ ($0<s<1$). It follows that $S^\dagger(s)qS(s)=sq$. Thus the measurement induces a squeezing on an arbitrary mode of any hyperedge connected to the measured vertex. Without loss of generality, let this be the $(k-1)$th vertex. Therefore,
\begin{widetext}
\begin{align}
\ket{m}_{q,j}\braket{m|g_{\le n}}&=\prod_k\prod_{e_k\in E^k_-}CZ_{e_k}\prod_{e_{k-1}\in E^{k-1}_+}F^\dagger_kZ_k(m)e^{imq_{i_1}\dots q_{i_{k-1}}}e^{iq_{i_1}\dots q_{i_{k-1}}p_k}\ket{0}_p^{\otimes n-1}\ket{0}_{p_k}\\
&=\prod_k\prod_{e_k\in E^k_-}CZ_{e_k}\prod_{e_{k-1}\in E^{k-1}_+}e^{imq_{i_1}\dots q_{i_{k-1}}}F^\dagger_kZ_k(m)\ket{0}_p^{\otimes n-1}\ket{0}_{p_k}\\
&=\prod_k\prod_{e_k\in E^k_-}CZ_{e_k}\prod_{e_{k-1}\in E^{k-1}_+}e^{imq_{i_1}\dots q_{i_{k-1}}}\ket{0}_p^{\otimes n-1}F^\dagger_kZ_k(m)\ket{0}_{p_k}\\
&=\prod_k\prod_{e_k\in E^k_-}CZ_{e_k}\prod_{e_{k-1}\in E^{k-1}_+}S^\dagger_{k-1}(m)CZ_{e_{k-1}}\ket{0}_p^{\otimes n-1}F^\dagger_kZ_k(m)\ket{0}_{p_k}\,.
\end{align}
\end{widetext}
This state is the hypergraph state $\ket{g_{\le n}}$ with the measured vertex disconnected from the graph while all hyperedges associated with the measured vertex suffer a decrease in order by one, as claimed. \qed 

 \textbf{Lemma 2}: \textit{A $p$ measurement on a vertex of a hypergraph state converts any hyperedges on the affected vertex from $k$-hyperedges to $(k-1)$-hyperedges The newly generated hyperedges are in a superposition of all possible weights, with a phase determined by the weight and the measurement result.}
 
\textit{Proof}: The proof proceeds in the same fashion as that of Lemma 1, the core element of which is the effect on hyperedges connected to the measured vertex. Consider the same notation and for simplicity restrict attention to the set of k-hyperedges $E_+^k$. Then, 
\begin{widetext}
\begin{align}
\ket{m}_{p,j}\braket{m|g_{\le n}}&=\prod_k\prod_{e_k\in E_+^k}\ket{m}_{p,j}\bra{m}e^{iq_{i_1}\dots q_{i_k}}\ket{0}_p^{\otimes n}\\
&=\prod_k\prod_{e_k\in E_+^k}\int dx\ket{m}_{p,j}\bra{m}\ket{x}_{q,j}\bra{x}e^{iq_{i_1}\dots q_{i_k}}\ket{0}_p^{\otimes n}\\
&=\prod_k\prod_{e_k\in E_+^k}\int dxe^{-imx}e^{ixq_{i_1}\dots q_{i_{k-1}}}\ket{0}_p^{\otimes n-1}\ket{m}_{p_j}
\end{align}
\end{widetext}
\qed

Observe that these new edges can be 1-edges which are displacements. These can be manipulated to sometimes present more interesting states. For example, in the case of a 3-vertex cluster state a measurement on the middle vertex creates a superposition that can be manipulated to show that the result is exactly a $CZ_{i,j}$ gate between the first and last vertices.

\bibliography{references}

 \newcommand{\noop}[1]{}
\begin{thebibliography}{50}%
\makeatletter
\providecommand \@ifxundefined [1]{%
 \@ifx{#1\undefined}
}%
\providecommand \@ifnum [1]{%
 \ifnum #1\expandafter \@firstoftwo
 \else \expandafter \@secondoftwo
 \fi
}%
\providecommand \@ifx [1]{%
 \ifx #1\expandafter \@firstoftwo
 \else \expandafter \@secondoftwo
 \fi
}%
\providecommand \natexlab [1]{#1}%
\providecommand \enquote  [1]{``#1''}%
\providecommand \bibnamefont  [1]{#1}%
\providecommand \bibfnamefont [1]{#1}%
\providecommand \citenamefont [1]{#1}%
\providecommand \href@noop [0]{\@secondoftwo}%
\providecommand \href [0]{\begingroup \@sanitize@url \@href}%
\providecommand \@href[1]{\@@startlink{#1}\@@href}%
\providecommand \@@href[1]{\endgroup#1\@@endlink}%
\providecommand \@sanitize@url [0]{\catcode `\\12\catcode `\$12\catcode
  `\&12\catcode `\#12\catcode `\^12\catcode `\_12\catcode `\%12\relax}%
\providecommand \@@startlink[1]{}%
\providecommand \@@endlink[0]{}%
\providecommand \url  [0]{\begingroup\@sanitize@url \@url }%
\providecommand \@url [1]{\endgroup\@href {#1}{\urlprefix }}%
\providecommand \urlprefix  [0]{URL }%
\providecommand \Eprint [0]{\href }%
\providecommand \doibase [0]{http://dx.doi.org/}%
\providecommand \selectlanguage [0]{\@gobble}%
\providecommand \bibinfo  [0]{\@secondoftwo}%
\providecommand \bibfield  [0]{\@secondoftwo}%
\providecommand \translation [1]{[#1]}%
\providecommand \BibitemOpen [0]{}%
\providecommand \bibitemStop [0]{}%
\providecommand \bibitemNoStop [0]{.\EOS\space}%
\providecommand \EOS [0]{\spacefactor3000\relax}%
\providecommand \BibitemShut  [1]{\csname bibitem#1\endcsname}%
\let\auto@bib@innerbib\@empty
%</preamble>
\bibitem [{\citenamefont {Thompson}\ \emph {et~al.}(2008)\citenamefont
  {Thompson}, \citenamefont {Zwickl}, \citenamefont {Jayich}, \citenamefont
  {Marquardt}, \citenamefont {Girvin},\ and\ \citenamefont
  {Harris}}]{thompson2008strong}%
  \BibitemOpen
  \bibfield  {author} {\bibinfo {author} {\bibfnamefont {J.~D.}\ \bibnamefont
  {Thompson}}, \bibinfo {author} {\bibfnamefont {B.~M.}\ \bibnamefont
  {Zwickl}}, \bibinfo {author} {\bibfnamefont {A.~M.}\ \bibnamefont {Jayich}},
  \bibinfo {author} {\bibfnamefont {F.}~\bibnamefont {Marquardt}}, \bibinfo
  {author} {\bibfnamefont {S.~M.}\ \bibnamefont {Girvin}}, \ and\ \bibinfo
  {author} {\bibfnamefont {J.~G.~E.}\ \bibnamefont {Harris}},\ }\href
  {https://www.nature.com/articles/nature06715} {\bibfield  {journal} {\bibinfo
   {journal} {Nature}\ }\textbf {\bibinfo {volume} {452}},\ \bibinfo {pages}
  {72} (\bibinfo {year} {2008})}\BibitemShut {NoStop}%
\bibitem [{\citenamefont {Karuza}\ \emph {et~al.}(2012)\citenamefont {Karuza},
  \citenamefont {Galassi}, \citenamefont {Biancofiore}, \citenamefont
  {Molinelli}, \citenamefont {Natali}, \citenamefont {Tombesi}, \citenamefont
  {Giuseppe},\ and\ \citenamefont {Vitali}}]{karuza2012tunable}%
  \BibitemOpen
  \bibfield  {author} {\bibinfo {author} {\bibfnamefont {M.}~\bibnamefont
  {Karuza}}, \bibinfo {author} {\bibfnamefont {M.}~\bibnamefont {Galassi}},
  \bibinfo {author} {\bibfnamefont {C.}~\bibnamefont {Biancofiore}}, \bibinfo
  {author} {\bibfnamefont {C.}~\bibnamefont {Molinelli}}, \bibinfo {author}
  {\bibfnamefont {R.}~\bibnamefont {Natali}}, \bibinfo {author} {\bibfnamefont
  {P.}~\bibnamefont {Tombesi}}, \bibinfo {author} {\bibfnamefont {G.~D.}\
  \bibnamefont {Giuseppe}}, \ and\ \bibinfo {author} {\bibfnamefont
  {D.}~\bibnamefont {Vitali}},\ }\href {\doibase
  https://doi.org/10.1088/2040-8978/15/2/025704} {\bibfield  {journal}
  {\bibinfo  {journal} {J. Opt.}\ }\textbf {\bibinfo {volume} {15}},\ \bibinfo
  {pages} {025704} (\bibinfo {year} {2012})}\BibitemShut {NoStop}%
\bibitem [{\citenamefont {Para{\" i}so}\ \emph {et~al.}(2015)\citenamefont
  {Para{\" i}so}, \citenamefont {Kalaee}, \citenamefont {Zang}, \citenamefont
  {Pfeifer}, \citenamefont {Marquardt},\ and\ \citenamefont
  {Painter}}]{paraiso2015positionsquared}%
  \BibitemOpen
  \bibfield  {author} {\bibinfo {author} {\bibfnamefont {T.~K.}\ \bibnamefont
  {Para{\" i}so}}, \bibinfo {author} {\bibfnamefont {M.}~\bibnamefont
  {Kalaee}}, \bibinfo {author} {\bibfnamefont {L.}~\bibnamefont {Zang}},
  \bibinfo {author} {\bibfnamefont {H.}~\bibnamefont {Pfeifer}}, \bibinfo
  {author} {\bibfnamefont {F.}~\bibnamefont {Marquardt}}, \ and\ \bibinfo
  {author} {\bibfnamefont {O.}~\bibnamefont {Painter}},\ }\href {\doibase
  https://doi.org/10.1103/PhysRevX.5.041024} {\bibfield  {journal} {\bibinfo
  {journal} {Phys. Rev. X}\ }\textbf {\bibinfo {volume} {5}},\ \bibinfo {pages}
  {041024} (\bibinfo {year} {2015})}\BibitemShut {NoStop}%
\bibitem [{\citenamefont {Kalaee}\ \emph {et~al.}(2016)\citenamefont {Kalaee},
  \citenamefont {Para{\" i}so}, \citenamefont {Pfeifer},\ and\ \citenamefont
  {Painter}}]{kalaee2016design}%
  \BibitemOpen
  \bibfield  {author} {\bibinfo {author} {\bibfnamefont {M.}~\bibnamefont
  {Kalaee}}, \bibinfo {author} {\bibfnamefont {T.~K.}\ \bibnamefont {Para{\"
  i}so}}, \bibinfo {author} {\bibfnamefont {H.}~\bibnamefont {Pfeifer}}, \ and\
  \bibinfo {author} {\bibfnamefont {O.}~\bibnamefont {Painter}},\ }\href
  {\doibase https://doi.org/10.1364/OE.24.021308} {\bibfield  {journal}
  {\bibinfo  {journal} {Optics Express}\ }\textbf {\bibinfo {volume} {24}},\
  \bibinfo {pages} {21308} (\bibinfo {year} {2016})}\BibitemShut {NoStop}%
\bibitem [{\citenamefont {Brawley}\ \emph {et~al.}(2016)\citenamefont
  {Brawley}, \citenamefont {Vanner}, \citenamefont {Larsen}, \citenamefont
  {Schmid}, \citenamefont {Bolsen},\ and\ \citenamefont
  {Bowen}}]{brawley2016nonlinear}%
  \BibitemOpen
  \bibfield  {author} {\bibinfo {author} {\bibfnamefont {G.~A.}\ \bibnamefont
  {Brawley}}, \bibinfo {author} {\bibfnamefont {M.~R.}\ \bibnamefont {Vanner}},
  \bibinfo {author} {\bibfnamefont {P.~E.}\ \bibnamefont {Larsen}}, \bibinfo
  {author} {\bibfnamefont {S.}~\bibnamefont {Schmid}}, \bibinfo {author}
  {\bibfnamefont {A.}~\bibnamefont {Bolsen}}, \ and\ \bibinfo {author}
  {\bibfnamefont {W.~P.}\ \bibnamefont {Bowen}},\ }\href
  {https://www.nature.com/articles/ncomms10988} {\bibfield  {journal} {\bibinfo
   {journal} {Nature Comm.}\ }\textbf {\bibinfo {volume} {7}},\ \bibinfo
  {pages} {10988} (\bibinfo {year} {2016})}\BibitemShut {NoStop}%
\bibitem [{\citenamefont {Gieseler}\ \emph {et~al.}(2013)\citenamefont
  {Gieseler}, \citenamefont {Novotny},\ and\ \citenamefont
  {Quidant}}]{gieseler2013thermal}%
  \BibitemOpen
  \bibfield  {author} {\bibinfo {author} {\bibfnamefont {J.}~\bibnamefont
  {Gieseler}}, \bibinfo {author} {\bibfnamefont {L.}~\bibnamefont {Novotny}}, \
  and\ \bibinfo {author} {\bibfnamefont {R.}~\bibnamefont {Quidant}},\ }\href
  {http://www.nature.com/articles/nphys2798} {\bibfield  {journal} {\bibinfo
  {journal} {Nature Phys.}\ }\textbf {\bibinfo {volume} {9}},\ \bibinfo {pages}
  {806} (\bibinfo {year} {2013})}\BibitemShut {NoStop}%
\bibitem [{\citenamefont {Kiesel}\ \emph {et~al.}(2014)\citenamefont {Kiesel},
  \citenamefont {Blaser}, \citenamefont {Deli\'{c}}, \citenamefont {Grass},
  \citenamefont {Kaltenbaek},\ and\ \citenamefont
  {Aspelmeyer}}]{kiesel2014cavity}%
  \BibitemOpen
  \bibfield  {author} {\bibinfo {author} {\bibfnamefont {N.}~\bibnamefont
  {Kiesel}}, \bibinfo {author} {\bibfnamefont {F.}~\bibnamefont {Blaser}},
  \bibinfo {author} {\bibfnamefont {U.}~\bibnamefont {Deli\'{c}}}, \bibinfo
  {author} {\bibfnamefont {D.}~\bibnamefont {Grass}}, \bibinfo {author}
  {\bibfnamefont {R.}~\bibnamefont {Kaltenbaek}}, \ and\ \bibinfo {author}
  {\bibfnamefont {M.}~\bibnamefont {Aspelmeyer}},\ }\href@noop {} {\bibfield
  {journal} {\bibinfo  {journal} {PNAS}\ }\textbf {\bibinfo {volume} {110}},\
  \bibinfo {pages} {14180} (\bibinfo {year} {2014})}\BibitemShut {NoStop}%
\bibitem [{\citenamefont {Fonseca}\ \emph {et~al.}(2016)\citenamefont
  {Fonseca}, \citenamefont {Aranas}, \citenamefont {Millen}, \citenamefont
  {Monteiro},\ and\ \citenamefont {Barker}}]{fonseca2016nonlinear}%
  \BibitemOpen
  \bibfield  {author} {\bibinfo {author} {\bibfnamefont {P.~Z.~G.}\
  \bibnamefont {Fonseca}}, \bibinfo {author} {\bibfnamefont {E.~B.}\
  \bibnamefont {Aranas}}, \bibinfo {author} {\bibfnamefont {J.}~\bibnamefont
  {Millen}}, \bibinfo {author} {\bibfnamefont {T.~S.}\ \bibnamefont
  {Monteiro}}, \ and\ \bibinfo {author} {\bibfnamefont {P.~F.}\ \bibnamefont
  {Barker}},\ }\href@noop {} {\bibfield  {journal} {\bibinfo  {journal} {Phys.
  Rev. Lett.}\ }\textbf {\bibinfo {volume} {117}},\ \bibinfo {pages} {173602}
  (\bibinfo {year} {2016})}\BibitemShut {NoStop}%
\bibitem [{\citenamefont {Rashid}\ \emph {et~al.}(2016)\citenamefont {Rashid},
  \citenamefont {Tufarelli}, \citenamefont {Bateman}, \citenamefont {Vovrosh},
  \citenamefont {Hempston}, \citenamefont {Kim},\ and\ \citenamefont
  {Ulbricht}}]{rashid2016experimental}%
  \BibitemOpen
  \bibfield  {author} {\bibinfo {author} {\bibfnamefont {M.}~\bibnamefont
  {Rashid}}, \bibinfo {author} {\bibfnamefont {T.}~\bibnamefont {Tufarelli}},
  \bibinfo {author} {\bibfnamefont {J.}~\bibnamefont {Bateman}}, \bibinfo
  {author} {\bibfnamefont {J.}~\bibnamefont {Vovrosh}}, \bibinfo {author}
  {\bibfnamefont {D.}~\bibnamefont {Hempston}}, \bibinfo {author}
  {\bibfnamefont {M.~S.}\ \bibnamefont {Kim}}, \ and\ \bibinfo {author}
  {\bibfnamefont {H.}~\bibnamefont {Ulbricht}},\ }\href
  {https://journals.aps.org/prl/abstract/10.1103/PhysRevLett.117.273601}
  {\bibfield  {journal} {\bibinfo  {journal} {Phys. Rev. Lett.}\ }\textbf
  {\bibinfo {volume} {117}},\ \bibinfo {pages} {273601} (\bibinfo {year}
  {2016})}\BibitemShut {NoStop}%
\bibitem [{\citenamefont {\v{S}iler}\ \emph {et~al.}(2017)\citenamefont
  {\v{S}iler}, \citenamefont {J\'{a}kl}, \citenamefont {Brzobohat\'{y}},
  \citenamefont {Ryabov}, \citenamefont {Filip},\ and\ \citenamefont
  {Zem\'{a}nek}}]{siler2017thermally}%
  \BibitemOpen
  \bibfield  {author} {\bibinfo {author} {\bibfnamefont {M.}~\bibnamefont
  {\v{S}iler}}, \bibinfo {author} {\bibfnamefont {P.}~\bibnamefont {J\'{a}kl}},
  \bibinfo {author} {\bibfnamefont {O.}~\bibnamefont {Brzobohat\'{y}}},
  \bibinfo {author} {\bibfnamefont {A.}~\bibnamefont {Ryabov}}, \bibinfo
  {author} {\bibfnamefont {R.}~\bibnamefont {Filip}}, \ and\ \bibinfo {author}
  {\bibfnamefont {P.}~\bibnamefont {Zem\'{a}nek}},\ }\href
  {https://www.nature.com/articles/s41598-017-01848-4#article-info} {\bibfield
  {journal} {\bibinfo  {journal} {Sci. Rep.}\ }\textbf {\bibinfo {volume}
  {7}},\ \bibinfo {pages} {1697} (\bibinfo {year} {2017})}\BibitemShut
  {NoStop}%
\bibitem [{\citenamefont {Ricci}\ \emph {et~al.}(2017)\citenamefont {Ricci},
  \citenamefont {Rica}, \citenamefont {Spasenovi\'{c}}, \citenamefont
  {Gieseler}, \citenamefont {Rondin}, \citenamefont {Novotny},\ and\
  \citenamefont {Quidant}}]{ricci2017optically}%
  \BibitemOpen
  \bibfield  {author} {\bibinfo {author} {\bibfnamefont {F.}~\bibnamefont
  {Ricci}}, \bibinfo {author} {\bibfnamefont {R.~A.}\ \bibnamefont {Rica}},
  \bibinfo {author} {\bibfnamefont {M.}~\bibnamefont {Spasenovi\'{c}}},
  \bibinfo {author} {\bibfnamefont {J.}~\bibnamefont {Gieseler}}, \bibinfo
  {author} {\bibfnamefont {L.}~\bibnamefont {Rondin}}, \bibinfo {author}
  {\bibfnamefont {L.}~\bibnamefont {Novotny}}, \ and\ \bibinfo {author}
  {\bibfnamefont {R.}~\bibnamefont {Quidant}},\ }\href
  {https://www.nature.com/articles/ncomms15141} {\bibfield  {journal} {\bibinfo
   {journal} {Nature Communications}\ }\textbf {\bibinfo {volume} {8}},\
  \bibinfo {pages} {15141} (\bibinfo {year} {2017})}\BibitemShut {NoStop}%
\bibitem [{\citenamefont {Ourjoumtsev}\ \emph {et~al.}(2007)\citenamefont
  {Ourjoumtsev}, \citenamefont {Dantan}, \citenamefont {Tualle-Brouri},\ and\
  \citenamefont {Grangier}}]{ourjoumtsev2007increasing}%
  \BibitemOpen
  \bibfield  {author} {\bibinfo {author} {\bibfnamefont {A.}~\bibnamefont
  {Ourjoumtsev}}, \bibinfo {author} {\bibfnamefont {A.}~\bibnamefont {Dantan}},
  \bibinfo {author} {\bibfnamefont {R.}~\bibnamefont {Tualle-Brouri}}, \ and\
  \bibinfo {author} {\bibfnamefont {P.}~\bibnamefont {Grangier}},\ }\href@noop
  {} {\bibfield  {journal} {\bibinfo  {journal} {Phys. Rev. Lett.}\ }\textbf
  {\bibinfo {volume} {98}},\ \bibinfo {pages} {030502} (\bibinfo {year}
  {2007})}\BibitemShut {NoStop}%
\bibitem [{\citenamefont {Marek}\ \emph {et~al.}(2012)\citenamefont {Marek},
  \citenamefont {Filip},\ and\ \citenamefont
  {Furusawa}}]{marek2012deterministic}%
  \BibitemOpen
  \bibfield  {author} {\bibinfo {author} {\bibfnamefont {P.}~\bibnamefont
  {Marek}}, \bibinfo {author} {\bibfnamefont {R.}~\bibnamefont {Filip}}, \ and\
  \bibinfo {author} {\bibfnamefont {A.}~\bibnamefont {Furusawa}},\ }\href
  {https://journals.aps.org/pra/abstract/10.1103/PhysRevA.84.053802} {\bibfield
   {journal} {\bibinfo  {journal} {Phys. Rev. A}\ }\textbf {\bibinfo {volume}
  {84}},\ \bibinfo {pages} {053802} (\bibinfo {year} {2012})}\BibitemShut
  {NoStop}%
\bibitem [{\citenamefont {Marshall}\ \emph {et~al.}(2015)\citenamefont
  {Marshall}, \citenamefont {Pooser}, \citenamefont {Siopsis},\ and\
  \citenamefont {Weedbrook}}]{marshall2015repeat}%
  \BibitemOpen
  \bibfield  {author} {\bibinfo {author} {\bibfnamefont {K.}~\bibnamefont
  {Marshall}}, \bibinfo {author} {\bibfnamefont {R.}~\bibnamefont {Pooser}},
  \bibinfo {author} {\bibfnamefont {G.}~\bibnamefont {Siopsis}}, \ and\
  \bibinfo {author} {\bibfnamefont {C.}~\bibnamefont {Weedbrook}},\ }\href@noop
  {} {\bibfield  {journal} {\bibinfo  {journal} {Phys. Rev. A}\ }\textbf
  {\bibinfo {volume} {91}},\ \bibinfo {pages} {032321} (\bibinfo {year}
  {2015})}\BibitemShut {NoStop}%
\bibitem [{\citenamefont {Miyata}\ \emph {et~al.}(2016)\citenamefont {Miyata},
  \citenamefont {Ogawa}, \citenamefont {Marek}, \citenamefont {Filip},
  \citenamefont {Yonezawa},\ and\ \citenamefont
  {Furusawa}}]{miyata2016implementation}%
  \BibitemOpen
  \bibfield  {author} {\bibinfo {author} {\bibfnamefont {K.}~\bibnamefont
  {Miyata}}, \bibinfo {author} {\bibfnamefont {H.}~\bibnamefont {Ogawa}},
  \bibinfo {author} {\bibfnamefont {P.}~\bibnamefont {Marek}}, \bibinfo
  {author} {\bibfnamefont {R.}~\bibnamefont {Filip}}, \bibinfo {author}
  {\bibfnamefont {H.}~\bibnamefont {Yonezawa}}, \ and\ \bibinfo {author}
  {\bibfnamefont {A.}~\bibnamefont {Furusawa}},\ }\href {\doibase
  10.1103/PhysRevA.93.022301} {\bibfield  {journal} {\bibinfo  {journal} {Phys.
  Rev. A}\ }\textbf {\bibinfo {volume} {93}},\ \bibinfo {pages} {022301}
  (\bibinfo {year} {2016})}\BibitemShut {NoStop}%
\bibitem [{\citenamefont {Lloyd}\ and\ \citenamefont
  {Braunstein}(2003)}]{lloyd2003quantum}%
  \BibitemOpen
  \bibfield  {author} {\bibinfo {author} {\bibfnamefont {S.}~\bibnamefont
  {Lloyd}}\ and\ \bibinfo {author} {\bibfnamefont {S.~L.}\ \bibnamefont
  {Braunstein}},\ }in\ \href@noop {} {\emph {\bibinfo {booktitle} {Quantum
  Information with Continuous Variables}}}\ (\bibinfo  {publisher} {Springer},\
  \bibinfo {year} {2003})\ pp.\ \bibinfo {pages} {9--17}\BibitemShut {NoStop}%
\bibitem [{\citenamefont {Menicucci}\ \emph {et~al.}(2006)\citenamefont
  {Menicucci}, \citenamefont {van Loock}, \citenamefont {Gu}, \citenamefont
  {Weedbrook}, \citenamefont {Ralph},\ and\ \citenamefont
  {Nielsen}}]{menicucci2006universal}%
  \BibitemOpen
  \bibfield  {author} {\bibinfo {author} {\bibfnamefont {N.~C.}\ \bibnamefont
  {Menicucci}}, \bibinfo {author} {\bibfnamefont {P.}~\bibnamefont {van
  Loock}}, \bibinfo {author} {\bibfnamefont {M.}~\bibnamefont {Gu}}, \bibinfo
  {author} {\bibfnamefont {C.}~\bibnamefont {Weedbrook}}, \bibinfo {author}
  {\bibfnamefont {T.~C.}\ \bibnamefont {Ralph}}, \ and\ \bibinfo {author}
  {\bibfnamefont {M.~A.}\ \bibnamefont {Nielsen}},\ }\href {\doibase
  10.1103/PhysRevLett.97.110501} {\bibfield  {journal} {\bibinfo  {journal}
  {Phys. Rev. Lett.}\ }\textbf {\bibinfo {volume} {97}},\ \bibinfo {pages}
  {110501} (\bibinfo {year} {2006})}\BibitemShut {NoStop}%
\bibitem [{\citenamefont {Weedbrook}\ \emph {et~al.}(2012)\citenamefont
  {Weedbrook}, \citenamefont {Pirandola}, \citenamefont {Garc\'{\i}a-Patr\'on},
  \citenamefont {Cerf}, \citenamefont {Ralph}, \citenamefont {Shapiro},\ and\
  \citenamefont {Lloyd}}]{weedbrook2012gaussian}%
  \BibitemOpen
  \bibfield  {author} {\bibinfo {author} {\bibfnamefont {C.}~\bibnamefont
  {Weedbrook}}, \bibinfo {author} {\bibfnamefont {S.}~\bibnamefont
  {Pirandola}}, \bibinfo {author} {\bibfnamefont {R.}~\bibnamefont
  {Garc\'{\i}a-Patr\'on}}, \bibinfo {author} {\bibfnamefont {N.~J.}\
  \bibnamefont {Cerf}}, \bibinfo {author} {\bibfnamefont {T.~C.}\ \bibnamefont
  {Ralph}}, \bibinfo {author} {\bibfnamefont {J.~H.}\ \bibnamefont {Shapiro}},
  \ and\ \bibinfo {author} {\bibfnamefont {S.}~\bibnamefont {Lloyd}},\ }\href
  {\doibase 10.1103/RevModPhys.84.621} {\bibfield  {journal} {\bibinfo
  {journal} {Rev. Mod. Phys.}\ }\textbf {\bibinfo {volume} {84}},\ \bibinfo
  {pages} {621} (\bibinfo {year} {2012})}\BibitemShut {NoStop}%
\bibitem [{\citenamefont {Fiur\'{a}\v{s}ek}(2002)}]{fiurasek2002gaussian}%
  \BibitemOpen
  \bibfield  {author} {\bibinfo {author} {\bibfnamefont {J.}~\bibnamefont
  {Fiur\'{a}\v{s}ek}},\ }\href@noop {} {\bibfield  {journal} {\bibinfo
  {journal} {Phys. Rev. Lett.}\ }\textbf {\bibinfo {volume} {89}},\ \bibinfo
  {pages} {137904} (\bibinfo {year} {2002})}\BibitemShut {NoStop}%
\bibitem [{\citenamefont {Gu}\ \emph {et~al.}(2009)\citenamefont {Gu},
  \citenamefont {Weedbrook}, \citenamefont {Menicucci}, \citenamefont {Ralph},\
  and\ \citenamefont {van Loock}}]{gu2009quantum}%
  \BibitemOpen
  \bibfield  {author} {\bibinfo {author} {\bibfnamefont {M.}~\bibnamefont
  {Gu}}, \bibinfo {author} {\bibfnamefont {C.}~\bibnamefont {Weedbrook}},
  \bibinfo {author} {\bibfnamefont {N.~C.}\ \bibnamefont {Menicucci}}, \bibinfo
  {author} {\bibfnamefont {T.~C.}\ \bibnamefont {Ralph}}, \ and\ \bibinfo
  {author} {\bibfnamefont {P.}~\bibnamefont {van Loock}},\ }\href@noop {}
  {\bibfield  {journal} {\bibinfo  {journal} {Phys. Rev. A}\ }\textbf {\bibinfo
  {volume} {79}},\ \bibinfo {pages} {062318} (\bibinfo {year}
  {2009})}\BibitemShut {NoStop}%
\bibitem [{\citenamefont {Niset}\ \emph {et~al.}(2009)\citenamefont {Niset},
  \citenamefont {Fiur\'{a}\v{s}ek},\ and\ \citenamefont
  {Cerf}}]{niset2009nogo}%
  \BibitemOpen
  \bibfield  {author} {\bibinfo {author} {\bibfnamefont {J.}~\bibnamefont
  {Niset}}, \bibinfo {author} {\bibfnamefont {J.}~\bibnamefont
  {Fiur\'{a}\v{s}ek}}, \ and\ \bibinfo {author} {\bibfnamefont {N.~J.}\
  \bibnamefont {Cerf}},\ }\href@noop {} {\bibfield  {journal} {\bibinfo
  {journal} {Phys. Rev. A}\ }\textbf {\bibinfo {volume} {102}},\ \bibinfo
  {pages} {120501} (\bibinfo {year} {2009})}\BibitemShut {NoStop}%
\bibitem [{\citenamefont {Arzani}\ \emph {et~al.}(2017)\citenamefont {Arzani},
  \citenamefont {Treps},\ and\ \citenamefont {Ferrini}}]{arzani2017polynomial}%
  \BibitemOpen
  \bibfield  {author} {\bibinfo {author} {\bibfnamefont {F.}~\bibnamefont
  {Arzani}}, \bibinfo {author} {\bibfnamefont {N.}~\bibnamefont {Treps}}, \
  and\ \bibinfo {author} {\bibfnamefont {G.}~\bibnamefont {Ferrini}},\ }\href
  {https://journals.aps.org/pra/abstract/10.1103/PhysRevA.95.052352} {\bibfield
   {journal} {\bibinfo  {journal} {Phys. Rev. A}\ }\textbf {\bibinfo {volume}
  {95}},\ \bibinfo {pages} {052352} (\bibinfo {year} {2017})}\BibitemShut
  {NoStop}%
\bibitem [{\citenamefont {Walschaers}\ \emph {et~al.}(2018)\citenamefont
  {Walschaers}, \citenamefont {Sarkar}, \citenamefont {Parigi},\ and\
  \citenamefont {Treps}}]{walschaers2018tailoring}%
  \BibitemOpen
  \bibfield  {author} {\bibinfo {author} {\bibfnamefont {M.}~\bibnamefont
  {Walschaers}}, \bibinfo {author} {\bibfnamefont {S.}~\bibnamefont {Sarkar}},
  \bibinfo {author} {\bibfnamefont {V.}~\bibnamefont {Parigi}}, \ and\ \bibinfo
  {author} {\bibfnamefont {N.}~\bibnamefont {Treps}},\ }\href
  {https://journals.aps.org/prl/abstract/10.1103/PhysRevLett.121.220501}
  {\bibfield  {journal} {\bibinfo  {journal} {Phys. Rev. Lett.}\ }\textbf
  {\bibinfo {volume} {121}},\ \bibinfo {pages} {220501} (\bibinfo {year}
  {2018})}\BibitemShut {NoStop}%
\bibitem [{\citenamefont {Braunstein}\ \emph {et~al.}(2001)\citenamefont
  {Braunstein}, \citenamefont {Cerf}, \citenamefont {Iblisdir}, \citenamefont
  {van Loock},\ and\ \citenamefont {Massar}}]{braunstein2001otimal}%
  \BibitemOpen
  \bibfield  {author} {\bibinfo {author} {\bibfnamefont {S.~L.}\ \bibnamefont
  {Braunstein}}, \bibinfo {author} {\bibfnamefont {N.~J.}\ \bibnamefont
  {Cerf}}, \bibinfo {author} {\bibfnamefont {S.}~\bibnamefont {Iblisdir}},
  \bibinfo {author} {\bibfnamefont {P.}~\bibnamefont {van Loock}}, \ and\
  \bibinfo {author} {\bibfnamefont {S.}~\bibnamefont {Massar}},\ }\href@noop {}
  {\bibfield  {journal} {\bibinfo  {journal} {Phys. Rev. Lett.}\ }\textbf
  {\bibinfo {volume} {86}},\ \bibinfo {pages} {4938} (\bibinfo {year}
  {2001})}\BibitemShut {NoStop}%
\bibitem [{\citenamefont {Eisert}\ \emph {et~al.}(2002)\citenamefont {Eisert},
  \citenamefont {Scheel},\ and\ \citenamefont {Plenio}}]{eisert2002distilling}%
  \BibitemOpen
  \bibfield  {author} {\bibinfo {author} {\bibfnamefont {J.}~\bibnamefont
  {Eisert}}, \bibinfo {author} {\bibfnamefont {S.}~\bibnamefont {Scheel}}, \
  and\ \bibinfo {author} {\bibfnamefont {M.~B.}\ \bibnamefont {Plenio}},\
  }\href@noop {} {\bibfield  {journal} {\bibinfo  {journal} {Phys. Rev. Lett.}\
  }\textbf {\bibinfo {volume} {89}},\ \bibinfo {pages} {137903} (\bibinfo
  {year} {2002})}\BibitemShut {NoStop}%
\bibitem [{\citenamefont {Giedke}\ and\ \citenamefont
  {Cirac}(2002)}]{giedke2002characterization}%
  \BibitemOpen
  \bibfield  {author} {\bibinfo {author} {\bibfnamefont {G.}~\bibnamefont
  {Giedke}}\ and\ \bibinfo {author} {\bibfnamefont {J.~I.}\ \bibnamefont
  {Cirac}},\ }\href@noop {} {\bibfield  {journal} {\bibinfo  {journal} {Phys.
  Rev. A}\ }\textbf {\bibinfo {volume} {66}},\ \bibinfo {pages} {032316}
  (\bibinfo {year} {2002})}\BibitemShut {NoStop}%
\bibitem [{\citenamefont {Olivares}\ \emph {et~al.}(2003)\citenamefont
  {Olivares}, \citenamefont {Paris},\ and\ \citenamefont
  {Bonifacio}}]{olivares2003teleportation}%
  \BibitemOpen
  \bibfield  {author} {\bibinfo {author} {\bibfnamefont {S.}~\bibnamefont
  {Olivares}}, \bibinfo {author} {\bibfnamefont {M.~G.~A.}\ \bibnamefont
  {Paris}}, \ and\ \bibinfo {author} {\bibfnamefont {R.}~\bibnamefont
  {Bonifacio}},\ }\href@noop {} {\bibfield  {journal} {\bibinfo  {journal}
  {Phys. Rev. A}\ }\textbf {\bibinfo {volume} {67}},\ \bibinfo {pages} {032314}
  (\bibinfo {year} {2003})}\BibitemShut {NoStop}%
\bibitem [{\citenamefont {Adesso}\ \emph {et~al.}(2009)\citenamefont {Adesso},
  \citenamefont {Dell'Anno}, \citenamefont {Siena}, \citenamefont
  {Illuminati},\ and\ \citenamefont {Souza}}]{adesso2009optimal}%
  \BibitemOpen
  \bibfield  {author} {\bibinfo {author} {\bibfnamefont {G.}~\bibnamefont
  {Adesso}}, \bibinfo {author} {\bibfnamefont {F.}~\bibnamefont {Dell'Anno}},
  \bibinfo {author} {\bibfnamefont {S.~D.}\ \bibnamefont {Siena}}, \bibinfo
  {author} {\bibfnamefont {F.}~\bibnamefont {Illuminati}}, \ and\ \bibinfo
  {author} {\bibfnamefont {L.~A.~M.}\ \bibnamefont {Souza}},\ }\href@noop {}
  {\bibfield  {journal} {\bibinfo  {journal} {Phys. Rev. A}\ }\textbf {\bibinfo
  {volume} {79}},\ \bibinfo {pages} {040305(R)} (\bibinfo {year}
  {2009})}\BibitemShut {NoStop}%
\bibitem [{\citenamefont {Paternostro}\ \emph {et~al.}(2009)\citenamefont
  {Paternostro}, \citenamefont {Jeong},\ and\ \citenamefont
  {Ralph}}]{paternostro2009violations}%
  \BibitemOpen
  \bibfield  {author} {\bibinfo {author} {\bibfnamefont {M.}~\bibnamefont
  {Paternostro}}, \bibinfo {author} {\bibfnamefont {H.}~\bibnamefont {Jeong}},
  \ and\ \bibinfo {author} {\bibfnamefont {T.~C.}\ \bibnamefont {Ralph}},\
  }\href@noop {} {\bibfield  {journal} {\bibinfo  {journal} {Phys. Rev. A}\
  }\textbf {\bibinfo {volume} {79}},\ \bibinfo {pages} {012101} (\bibinfo
  {year} {2009})}\BibitemShut {NoStop}%
\bibitem [{\citenamefont {Magnin}\ \emph {et~al.}(2010)\citenamefont {Magnin},
  \citenamefont {Magniez}, \citenamefont {Leverrier},\ and\ \citenamefont
  {Cerf}}]{magnin2010strong}%
  \BibitemOpen
  \bibfield  {author} {\bibinfo {author} {\bibfnamefont {L.}~\bibnamefont
  {Magnin}}, \bibinfo {author} {\bibfnamefont {F.}~\bibnamefont {Magniez}},
  \bibinfo {author} {\bibfnamefont {A.}~\bibnamefont {Leverrier}}, \ and\
  \bibinfo {author} {\bibfnamefont {N.~J.}\ \bibnamefont {Cerf}},\ }\href@noop
  {} {\bibfield  {journal} {\bibinfo  {journal} {Phys. Rev. A}\ }\textbf
  {\bibinfo {volume} {81}},\ \bibinfo {pages} {010302(R)} (\bibinfo {year}
  {2010})}\BibitemShut {NoStop}%
\bibitem [{\citenamefont {Jabbour}\ \emph {et~al.}(2015)\citenamefont
  {Jabbour}, \citenamefont {Garc\'{i}a-Patr\'{o}n},\ and\ \citenamefont
  {Cerf}}]{jabbour2015interconversion}%
  \BibitemOpen
  \bibfield  {author} {\bibinfo {author} {\bibfnamefont {M.~G.}\ \bibnamefont
  {Jabbour}}, \bibinfo {author} {\bibfnamefont {R.}~\bibnamefont
  {Garc\'{i}a-Patr\'{o}n}}, \ and\ \bibinfo {author} {\bibfnamefont {N.~J.}\
  \bibnamefont {Cerf}},\ }\href@noop {} {\bibfield  {journal} {\bibinfo
  {journal} {Phys. Rev. A}\ }\textbf {\bibinfo {volume} {91}},\ \bibinfo
  {pages} {012316} (\bibinfo {year} {2015})}\BibitemShut {NoStop}%
\bibitem [{\citenamefont {Zhuang}\ \emph {et~al.}(2018)\citenamefont {Zhuang},
  \citenamefont {Shor},\ and\ \citenamefont {Shapiro}}]{zhuang2018resource}%
  \BibitemOpen
  \bibfield  {author} {\bibinfo {author} {\bibfnamefont {Q.}~\bibnamefont
  {Zhuang}}, \bibinfo {author} {\bibfnamefont {P.~W.}\ \bibnamefont {Shor}}, \
  and\ \bibinfo {author} {\bibfnamefont {J.~H.}\ \bibnamefont {Shapiro}},\
  }\href {https://journals.aps.org/pra/abstract/10.1103/PhysRevA.97.052317}
  {\bibfield  {journal} {\bibinfo  {journal} {Phys. Rev. A}\ }\textbf {\bibinfo
  {volume} {97}},\ \bibinfo {pages} {052317} (\bibinfo {year}
  {2018})}\BibitemShut {NoStop}%
\bibitem [{\citenamefont {Albarelli}\ \emph {et~al.}(2018)\citenamefont
  {Albarelli}, \citenamefont {Genoni}, \citenamefont {Paris},\ and\
  \citenamefont {Ferraro}}]{albarelli2018resource}%
  \BibitemOpen
  \bibfield  {author} {\bibinfo {author} {\bibfnamefont {F.}~\bibnamefont
  {Albarelli}}, \bibinfo {author} {\bibfnamefont {M.~G.}\ \bibnamefont
  {Genoni}}, \bibinfo {author} {\bibfnamefont {M.~G.~A.}\ \bibnamefont
  {Paris}}, \ and\ \bibinfo {author} {\bibfnamefont {A.}~\bibnamefont
  {Ferraro}},\ }\href
  {https://journals.aps.org/pra/abstract/10.1103/PhysRevA.98.052350} {\bibfield
   {journal} {\bibinfo  {journal} {Phys. Rev. A}\ }\textbf {\bibinfo {volume}
  {98}},\ \bibinfo {pages} {052350} (\bibinfo {year} {2018})}\BibitemShut
  {NoStop}%
\bibitem [{\citenamefont {Miwa}\ \emph {et~al.}(2009)\citenamefont {Miwa},
  \citenamefont {Yoshikawa}, \citenamefont {van Loock},\ and\ \citenamefont
  {Furusawa}}]{miwa2009demonstration}%
  \BibitemOpen
  \bibfield  {author} {\bibinfo {author} {\bibfnamefont {Y.}~\bibnamefont
  {Miwa}}, \bibinfo {author} {\bibfnamefont {J.-i.}\ \bibnamefont {Yoshikawa}},
  \bibinfo {author} {\bibfnamefont {P.}~\bibnamefont {van Loock}}, \ and\
  \bibinfo {author} {\bibfnamefont {A.}~\bibnamefont {Furusawa}},\ }\href
  {\doibase 10.1103/PhysRevA.80.050303} {\bibfield  {journal} {\bibinfo
  {journal} {Phys. Rev. A}\ }\textbf {\bibinfo {volume} {80}},\ \bibinfo
  {pages} {050303} (\bibinfo {year} {2009})}\BibitemShut {NoStop}%
\bibitem [{\citenamefont {Ukai}\ \emph {et~al.}(2011)\citenamefont {Ukai},
  \citenamefont {Iwata}, \citenamefont {Shimokawa}, \citenamefont {Armstrong},
  \citenamefont {Politi}, \citenamefont {Yoshikawa}, \citenamefont {van
  Loock},\ and\ \citenamefont {Furusawa}}]{ukai2011demonstration}%
  \BibitemOpen
  \bibfield  {author} {\bibinfo {author} {\bibfnamefont {R.}~\bibnamefont
  {Ukai}}, \bibinfo {author} {\bibfnamefont {N.}~\bibnamefont {Iwata}},
  \bibinfo {author} {\bibfnamefont {Y.}~\bibnamefont {Shimokawa}}, \bibinfo
  {author} {\bibfnamefont {S.~C.}\ \bibnamefont {Armstrong}}, \bibinfo {author}
  {\bibfnamefont {A.}~\bibnamefont {Politi}}, \bibinfo {author} {\bibfnamefont
  {J.-I.}\ \bibnamefont {Yoshikawa}}, \bibinfo {author} {\bibfnamefont
  {P.}~\bibnamefont {van Loock}}, \ and\ \bibinfo {author} {\bibfnamefont
  {A.}~\bibnamefont {Furusawa}},\ }\href@noop {} {\bibfield  {journal}
  {\bibinfo  {journal} {Phys. Rev. Lett.}\ }\textbf {\bibinfo {volume} {106}},\
  \bibinfo {pages} {240504} (\bibinfo {year} {2011})}\BibitemShut {NoStop}%
\bibitem [{\citenamefont {Yokoyama}\ \emph {et~al.}(2013)\citenamefont
  {Yokoyama}, \citenamefont {Ukai}, \citenamefont {Armstrong}, \citenamefont
  {Sornphiphatphong}, \citenamefont {Kaji}, \citenamefont {Suzuki},
  \citenamefont {Yoshikawa}, \citenamefont {Yonezawa}, \citenamefont
  {Menicucci},\ and\ \citenamefont {Furusawa}}]{yokoyama2013ultra}%
  \BibitemOpen
  \bibfield  {author} {\bibinfo {author} {\bibfnamefont {S.}~\bibnamefont
  {Yokoyama}}, \bibinfo {author} {\bibfnamefont {R.}~\bibnamefont {Ukai}},
  \bibinfo {author} {\bibfnamefont {S.~C.}\ \bibnamefont {Armstrong}}, \bibinfo
  {author} {\bibfnamefont {C.}~\bibnamefont {Sornphiphatphong}}, \bibinfo
  {author} {\bibfnamefont {T.}~\bibnamefont {Kaji}}, \bibinfo {author}
  {\bibfnamefont {S.}~\bibnamefont {Suzuki}}, \bibinfo {author} {\bibfnamefont
  {J.-I.}\ \bibnamefont {Yoshikawa}}, \bibinfo {author} {\bibfnamefont
  {H.}~\bibnamefont {Yonezawa}}, \bibinfo {author} {\bibfnamefont {N.~C.}\
  \bibnamefont {Menicucci}}, \ and\ \bibinfo {author} {\bibfnamefont
  {A.}~\bibnamefont {Furusawa}},\ }\href@noop {} {\bibfield  {journal}
  {\bibinfo  {journal} {Nature Photonics}\ }\textbf {\bibinfo {volume} {7}},\
  \bibinfo {pages} {982} (\bibinfo {year} {2013})}\BibitemShut {NoStop}%
\bibitem [{\citenamefont {Yoshikawa}\ \emph {et~al.}(2016)\citenamefont
  {Yoshikawa}, \citenamefont {Yokoyama}, \citenamefont {Kaji}, \citenamefont
  {Sornphiphatphong}, \citenamefont {Shiozawa}, \citenamefont {Makino},\ and\
  \citenamefont {Furusawa}}]{yoshikawa2016invited}%
  \BibitemOpen
  \bibfield  {author} {\bibinfo {author} {\bibfnamefont {J.-i.}\ \bibnamefont
  {Yoshikawa}}, \bibinfo {author} {\bibfnamefont {S.}~\bibnamefont {Yokoyama}},
  \bibinfo {author} {\bibfnamefont {T.}~\bibnamefont {Kaji}}, \bibinfo {author}
  {\bibfnamefont {C.}~\bibnamefont {Sornphiphatphong}}, \bibinfo {author}
  {\bibfnamefont {Y.}~\bibnamefont {Shiozawa}}, \bibinfo {author}
  {\bibfnamefont {K.}~\bibnamefont {Makino}}, \ and\ \bibinfo {author}
  {\bibfnamefont {A.}~\bibnamefont {Furusawa}},\ }\href@noop {} {\bibfield
  {journal} {\bibinfo  {journal} {APL Photonics}\ }\textbf {\bibinfo {volume}
  {1}},\ \bibinfo {pages} {060801} (\bibinfo {year} {2016})}\BibitemShut
  {NoStop}%
\bibitem [{\citenamefont {Bravyi}\ and\ \citenamefont
  {Kitaev}(2005)}]{bravyi2005universal}%
  \BibitemOpen
  \bibfield  {author} {\bibinfo {author} {\bibfnamefont {S.}~\bibnamefont
  {Bravyi}}\ and\ \bibinfo {author} {\bibfnamefont {A.}~\bibnamefont
  {Kitaev}},\ }\href {\doibase 10.1103/PhysRevA.71.022316} {\bibfield
  {journal} {\bibinfo  {journal} {Phys. Rev. A}\ }\textbf {\bibinfo {volume}
  {71}},\ \bibinfo {pages} {022316} (\bibinfo {year} {2005})}\BibitemShut
  {NoStop}%
\bibitem [{\citenamefont {Menicucci}(2014)}]{menicucci2014fault}%
  \BibitemOpen
  \bibfield  {author} {\bibinfo {author} {\bibfnamefont {N.~C.}\ \bibnamefont
  {Menicucci}},\ }\href {\doibase 10.1103/PhysRevLett.112.120504} {\bibfield
  {journal} {\bibinfo  {journal} {Phys. Rev. Lett.}\ }\textbf {\bibinfo
  {volume} {112}},\ \bibinfo {pages} {120504} (\bibinfo {year}
  {2014})}\BibitemShut {NoStop}%
\bibitem [{\citenamefont {Houhou}\ \emph {et~al.}(2018)\citenamefont {Houhou},
  \citenamefont {Moore}, \citenamefont {Bose},\ and\ \citenamefont
  {Ferraro}}]{houhou2018unconditional}%
  \BibitemOpen
  \bibfield  {author} {\bibinfo {author} {\bibfnamefont {O.}~\bibnamefont
  {Houhou}}, \bibinfo {author} {\bibfnamefont {D.~W.}\ \bibnamefont {Moore}},
  \bibinfo {author} {\bibfnamefont {S.}~\bibnamefont {Bose}}, \ and\ \bibinfo
  {author} {\bibfnamefont {A.}~\bibnamefont {Ferraro}},\ }\href
  {https://arxiv.org/abs/1809.09733} {\bibfield  {journal} {\bibinfo  {journal}
  {arxiv:1809.09733}\ } (\bibinfo {year} {2018})}\BibitemShut {NoStop}%
\bibitem [{\citenamefont {Gottesman}\ \emph {et~al.}(2001)\citenamefont
  {Gottesman}, \citenamefont {Kitaev},\ and\ \citenamefont
  {Preskill}}]{gottesman2001encoding}%
  \BibitemOpen
  \bibfield  {author} {\bibinfo {author} {\bibfnamefont {D.}~\bibnamefont
  {Gottesman}}, \bibinfo {author} {\bibfnamefont {A.}~\bibnamefont {Kitaev}}, \
  and\ \bibinfo {author} {\bibfnamefont {J.}~\bibnamefont {Preskill}},\ }\href
  {\doibase 10.1103/PhysRevA.64.012310} {\bibfield  {journal} {\bibinfo
  {journal} {Phys. Rev. A}\ }\textbf {\bibinfo {volume} {64}},\ \bibinfo
  {pages} {012310} (\bibinfo {year} {2001})}\BibitemShut {NoStop}%
\bibitem [{\citenamefont {Baragiola}\ \emph {et~al.}(2019)\citenamefont
  {Baragiola}, \citenamefont {Pantaleoni}, \citenamefont {Alexander},
  \citenamefont {Karanjai},\ and\ \citenamefont
  {Menicucci}}]{baragiola2019all}%
  \BibitemOpen
  \bibfield  {author} {\bibinfo {author} {\bibfnamefont {B.~Q.}\ \bibnamefont
  {Baragiola}}, \bibinfo {author} {\bibfnamefont {G.}~\bibnamefont
  {Pantaleoni}}, \bibinfo {author} {\bibfnamefont {R.~N.}\ \bibnamefont
  {Alexander}}, \bibinfo {author} {\bibfnamefont {A.}~\bibnamefont {Karanjai}},
  \ and\ \bibinfo {author} {\bibfnamefont {N.~C.}\ \bibnamefont {Menicucci}},\
  }\href {https://arxiv.org/abs/1903.00012} {\bibfield  {journal} {\bibinfo
  {journal} {arxiv:1903.00012}\ } (\bibinfo {year} {2019})}\BibitemShut
  {NoStop}%
\bibitem [{\citenamefont {Rossi}\ \emph {et~al.}(2013)\citenamefont {Rossi},
  \citenamefont {Huber}, \citenamefont {Bru\ss},\ and\ \citenamefont
  {Macchiavello}}]{rossi2013quantum}%
  \BibitemOpen
  \bibfield  {author} {\bibinfo {author} {\bibfnamefont {M.}~\bibnamefont
  {Rossi}}, \bibinfo {author} {\bibfnamefont {M.}~\bibnamefont {Huber}},
  \bibinfo {author} {\bibfnamefont {D.}~\bibnamefont {Bru\ss}}, \ and\ \bibinfo
  {author} {\bibfnamefont {C.}~\bibnamefont {Macchiavello}},\ }\href
  {https://iopscience.iop.org/article/10.1088/1367-2630/15/11/113022/meta}
  {\bibfield  {journal} {\bibinfo  {journal} {New J. Phys.}\ }\textbf {\bibinfo
  {volume} {15}},\ \bibinfo {pages} {113022} (\bibinfo {year}
  {2013})}\BibitemShut {NoStop}%
\bibitem [{\citenamefont {Qu}\ \emph {et~al.}(2013)\citenamefont {Qu},
  \citenamefont {Wang}, \citenamefont {Li},\ and\ \citenamefont
  {Bao}}]{qu2013encoding}%
  \BibitemOpen
  \bibfield  {author} {\bibinfo {author} {\bibfnamefont {R.}~\bibnamefont
  {Qu}}, \bibinfo {author} {\bibfnamefont {J.}~\bibnamefont {Wang}}, \bibinfo
  {author} {\bibfnamefont {Z.-S.}\ \bibnamefont {Li}}, \ and\ \bibinfo {author}
  {\bibfnamefont {Y.-R.}\ \bibnamefont {Bao}},\ }\href
  {https://journals.aps.org/pra/abstract/10.1103/PhysRevA.87.022311} {\bibfield
   {journal} {\bibinfo  {journal} {Phys. Rev. A}\ }\textbf {\bibinfo {volume}
  {87}},\ \bibinfo {pages} {022311} (\bibinfo {year} {2013})}\BibitemShut
  {NoStop}%
\bibitem [{\citenamefont {Miller}\ and\ \citenamefont
  {Miyake}(2016)}]{miller2016hierarchy}%
  \BibitemOpen
  \bibfield  {author} {\bibinfo {author} {\bibfnamefont {J.}~\bibnamefont
  {Miller}}\ and\ \bibinfo {author} {\bibfnamefont {A.}~\bibnamefont
  {Miyake}},\ }\href {https://www.nature.com/articles/npjqi201636} {\bibfield
  {journal} {\bibinfo  {journal} {npj Quantum Information}\ }\textbf {\bibinfo
  {volume} {2}},\ \bibinfo {pages} {16036} (\bibinfo {year}
  {2016})}\BibitemShut {NoStop}%
\bibitem [{\citenamefont {Devakul}\ and\ \citenamefont
  {Williamson}(2018)}]{devakul2018universal}%
  \BibitemOpen
  \bibfield  {author} {\bibinfo {author} {\bibfnamefont {T.}~\bibnamefont
  {Devakul}}\ and\ \bibinfo {author} {\bibfnamefont {D.~J.}\ \bibnamefont
  {Williamson}},\ }\href
  {https://journals.aps.org/pra/abstract/10.1103/PhysRevA.98.022332} {\bibfield
   {journal} {\bibinfo  {journal} {Phys. Rev. A}\ }\textbf {\bibinfo {volume}
  {98}},\ \bibinfo {pages} {022332} (\bibinfo {year} {2018})}\BibitemShut
  {NoStop}%
\bibitem [{\citenamefont {Miller}\ and\ \citenamefont
  {Miyake}(2018)}]{miller2018latent}%
  \BibitemOpen
  \bibfield  {author} {\bibinfo {author} {\bibfnamefont {J.}~\bibnamefont
  {Miller}}\ and\ \bibinfo {author} {\bibfnamefont {A.}~\bibnamefont
  {Miyake}},\ }\href
  {https://journals.aps.org/prl/abstract/10.1103/PhysRevLett.120.170503}
  {\bibfield  {journal} {\bibinfo  {journal} {Phys. Rev. Lett.}\ }\textbf
  {\bibinfo {volume} {120}},\ \bibinfo {pages} {170503} (\bibinfo {year}
  {2018})}\BibitemShut {NoStop}%
\bibitem [{\citenamefont {Kissinger}\ and\ \citenamefont {van~de
  Wetering}(2019)}]{kissinger2019universal}%
  \BibitemOpen
  \bibfield  {author} {\bibinfo {author} {\bibfnamefont {A.}~\bibnamefont
  {Kissinger}}\ and\ \bibinfo {author} {\bibfnamefont {J.}~\bibnamefont {van~de
  Wetering}},\ }\href
  {https://quantum-journal.org/papers/q-2019-04-26-134/?fbclid=IwAR2TgPrwXmpjlySUU-VCbX807Cap8W3IqETsfQsMlvttIlHnKaK-_hxQJ4s}
  {\bibfield  {journal} {\bibinfo  {journal} {Quantum}\ }\textbf {\bibinfo
  {volume} {3}},\ \bibinfo {pages} {134} (\bibinfo {year} {2019})}\BibitemShut
  {NoStop}%
\bibitem [{\citenamefont {Frattini}\ \emph {et~al.}(2017)\citenamefont
  {Frattini}, \citenamefont {Vool}, \citenamefont {Shankar}, \citenamefont
  {Narla}, \citenamefont {Sliwa},\ and\ \citenamefont
  {Devoret}}]{frattini20173wave}%
  \BibitemOpen
  \bibfield  {author} {\bibinfo {author} {\bibfnamefont {N.~E.}\ \bibnamefont
  {Frattini}}, \bibinfo {author} {\bibfnamefont {U.}~\bibnamefont {Vool}},
  \bibinfo {author} {\bibfnamefont {S.}~\bibnamefont {Shankar}}, \bibinfo
  {author} {\bibfnamefont {A.}~\bibnamefont {Narla}}, \bibinfo {author}
  {\bibfnamefont {K.~M.}\ \bibnamefont {Sliwa}}, \ and\ \bibinfo {author}
  {\bibfnamefont {M.~H.}\ \bibnamefont {Devoret}},\ }\href
  {https://aip.scitation.org/doi/10.1063/1.4984142} {\bibfield  {journal}
  {\bibinfo  {journal} {Appl. Phys. Lett.}\ }\textbf {\bibinfo {volume}
  {110}},\ \bibinfo {pages} {222603} (\bibinfo {year} {2017})}\BibitemShut
  {NoStop}%
\bibitem [{\citenamefont {Moore}\ \emph {et~al.}(2019)\citenamefont {Moore},
  \citenamefont {Rakhubovsky},\ and\ \citenamefont
  {Filip}}]{moore2019estimation}%
  \BibitemOpen
  \bibfield  {author} {\bibinfo {author} {\bibfnamefont {D.}~\bibnamefont
  {Moore}}, \bibinfo {author} {\bibfnamefont {A.~A.}\ \bibnamefont
  {Rakhubovsky}}, \ and\ \bibinfo {author} {\bibfnamefont {R.}~\bibnamefont
  {Filip}},\ }\href {https://arxiv.org/abs/1904.00773} {\bibfield  {journal}
  {\bibinfo  {journal} {arxiv:1904.00773}\ } (\bibinfo {year}
  {2019})}\BibitemShut {NoStop}%
\end{thebibliography}%

\end{document}